\documentclass[letterpaper,rmp,twocolumn,amsmath,amssymb,superscriptaddress,floatfix]{revtex4}
\usepackage{graphicx}
\RequirePackage[sort&compress]{natbib}
\usepackage{natbib}
\usepackage{color}
\usepackage{textcomp}
\usepackage{makecell}
\usepackage{mathptmx}
\usepackage{float}
\usepackage{setspace}
\usepackage[a]{esvect}
\usepackage[font=sf,size=small,justification=RaggedRight,singlelinecheck=false]{caption}
\DeclareCaptionLabelSeparator{bar}{ \rule[-0.2em]{0.3ex}{1em} }
\usepackage[sf,bf,small]{titlesec}
\titlespacing*{\section}{0pt}{1em}{0em}

\definecolor{darkgray}{rgb}{0.25,0.25,0.25}
\definecolor{darkred}{rgb}{0.89,0.10,0.11}
\definecolor{darkblue}{rgb}{0.12,0.39,0.62}

\usepackage[breaklinks=true,colorlinks=true,citecolor=black,linkcolor=black,menucolor=black,urlcolor=darkblue,pdfborder={1 0 0}]{hyperref}
\hypersetup{pdftitle={Mapping flows on weighted and directed networks with incomplete observations},pdfauthor={J. Smiljani\'c, C, Bl\"ocker, D. Edler, and M. Rosvall (2021)}}
\bibpunct{}{}{,}{s}{,}{,}

\usepackage{booktabs}

\newcommand{\mytoprule}{\specialrule{0.1em}{0em}{0em}}
\newcommand{\mybottomrule}{\specialrule{0.1em}{0em}{0em}} 
\newcommand{\mymidrule}{\specialrule{0.05em}{0em}{0em}}

\usepackage{dcolumn}
\usepackage{bm}
\usepackage{array} \newsavebox\cellbox
\newcolumntype{W}[2]
{>{\begin{lrbox}\cellbox}%
l%
<{\end{lrbox}%
\makebox[#2][#1]{\usebox\cellbox}}}
\usepackage{multirow}

\usepackage{cancel}	

\usepackage{amsthm,scrextend}
\usepackage{fancyhdr}

\begin{document}
\makeatletter
\renewcommand\@biblabel[1]{#1.}
\makeatother
	
\renewcommand{\figurename}{Figure}
\renewcommand{\thefigure}{\arabic{figure}}
\renewcommand{\tablename}{Table}
\renewcommand{\thetable}{\arabic{table}}
\renewcommand{\refname}{\large References}

\addtolength{\textheight}{1cm}
\addtolength{\textwidth}{1cm}
\addtolength{\hoffset}{-0.5cm}

\setlength{\belowcaptionskip}{1ex}
\setlength{\textfloatsep}{2ex}
\setlength{\dbltextfloatsep}{2ex}

\hyphenation{page-rank}

\newcommand*{\citen}[1]{%
  \begingroup
    \romannumeral-`\x 
    \setcitestyle{numbers}%
    \cite{#1}%
  \endgroup   
}

\newcommand{\enter}{\curvearrowleft}
\newcommand{\exit}{\curvearrowright}
\newcommand{\intra}{\circlearrowright}

\title{Mapping flows on weighted and directed networks with incomplete observations}

\author{Jelena Smiljani\'c}
\email{jelena.smiljanic@umu.se}
\affiliation{Integrated Science Lab, Department of Physics, Ume{\aa} University, SE-901 87 Ume{\aa}, Sweden}
\affiliation{Scientific Computing Laboratory, Center for the Study of Complex Systems, Institute of Physics Belgrade, University of Belgrade, Pregrevica 118, 11080 Belgrade, Serbia}
\author{Christopher Bl\"ocker}
\affiliation{Integrated Science Lab, Department of Physics, Ume{\aa} University, SE-901 87 Ume{\aa}, Sweden}
\author{Daniel Edler}
\affiliation{Integrated Science Lab, Department of Physics, Ume{\aa} University, SE-901 87 Ume{\aa}, Sweden}
\affiliation{Gothenburg Global Biodiversity Centre, Box 461, SE-405 30 Gothenburg, Sweden}
\affiliation{Department of Biological and Environmental Sciences, University of Gothenburg, Carl Skottsbergs gata 22B, Gothenburg 41319, Sweden}
\author{Martin Rosvall}
\affiliation{Integrated Science Lab, Department of Physics, Ume{\aa} University, SE-901 87 Ume{\aa}, Sweden}


\begin{abstract}
Detecting significant community structure in networks with incomplete observations is challenging because the evidence for specific solutions fades away with missing data.
For example, recent research shows that flow-based community detection methods can highlight spurious communities in sparse undirected and unweighted networks with missing links.
Current Bayesian approaches developed to overcome this problem do not work for incomplete observations in weighted and directed networks that describe network flows.
To overcome this gap, we extend the idea behind the Bayesian estimate of the map equation for unweighted and undirected networks to enable more robust community detection in weighted and directed networks.
We derive an empirical Bayes estimate of the transitions rates that can incorporate metadata information and show how an efficient implementation in the community-detection method Infomap provides more reliable communities even with a significant fraction of data missing.
\end{abstract}

\maketitle

\section{Introduction}

Network models gain explainable power with additional information about node labels or link directions and weights \cite{barrat2004architecture,newman2004analysis}.
But these data can also introduce uncertainties such as mislabeled nodes or noisy link measurements that the network methods must address for reliable further analysis \cite{newman2018network}.
For example, when community-detection methods disregard uncertainties in network data, they can overfit and generate inaccurate node classifications that affect downstream analyses such as link prediction \cite{ghasemian2019overfitting, ghasemian2020pnas, smiljanic2020pre}.

To assess the significance of detected communities, we can statistically compare them with expected results under a null model \cite{lancichinetti2010pre,lancichinetti2011plosone} or test how robust they are under random perturbations of the network \cite{rosvall2010plosone}.
However, both approaches are computationally expensive and impractical for large networks.
Instead, we can integrate regularization mechanisms in the community-detection methods themselves to prevent them from capitalizing on spurious communities. 
Several community detection methods take this approach for undirected and unweighted networks.
For example, community-detection methods based on statistical inference can incorporate  assumptions about unreliable measurements into the generative network models \cite{martin2016reconstruction, peixoto2018reconstruction}. 
For the flow-based community-detection method known as the map equation, which identifies modular structure by searching for sets of nodes with long flow persistence \cite{rosvall2008pnas, edler2017algorithms}, we have derived a Bayesian estimate that copes with missing unweighted and undirected links \cite{smiljanic2020pre}.
However, dealing with incomplete observations for robust flow-based community detection in directed and weighted networks remains unresolved. 

Since link weights and directions naturally describe network flows, the map equation works effectively for directed and weighted networks.
But the Bayesian estimate of the map equation for unweighted and undirected links requires an analytical expression for the network-flow distribution.
For directed networks, no such analytical solution exists.
Because the Bayesian estimate of the map equation also assumes a binary network to derive link probabilities, it cannot be applied directly to weighted and directed networks.

\begin{figure*}[ht]
    \includegraphics[width=\textwidth]{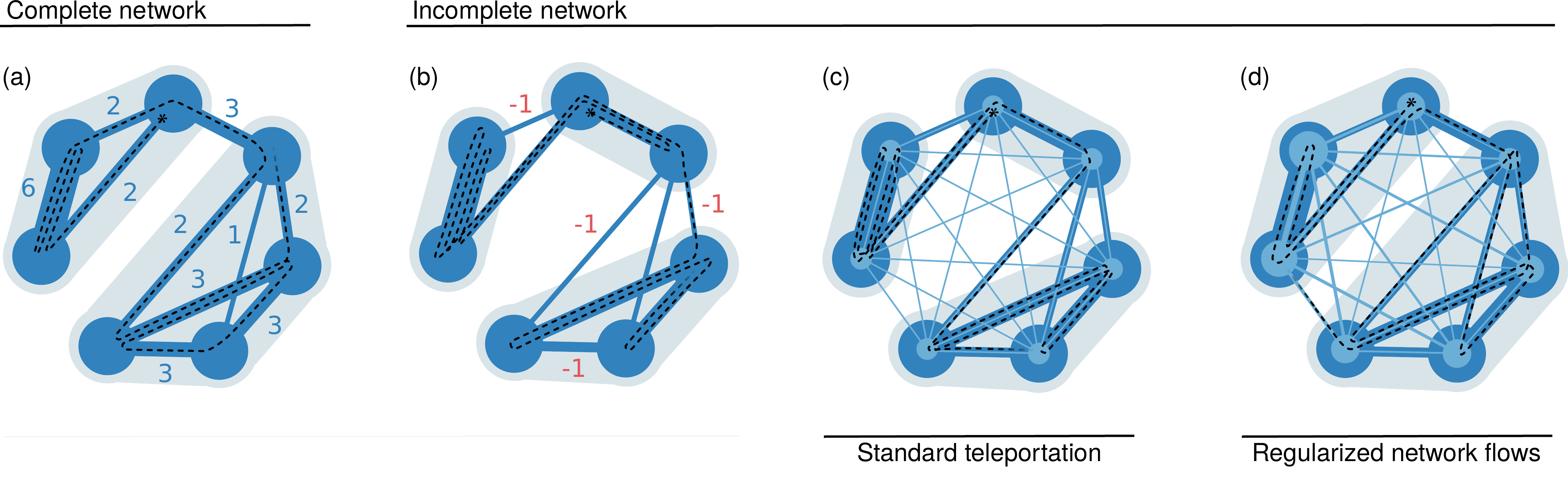}
    \caption{A schematic weighted network with complete and missing link observations. (a) A complete network with accurate network flows and inferred communities. (b) Missing link observations introduce inaccuracies. (c) A standard teleportation scheme cannot overcome the inaccuracies. (d) Regularized network flows with an empirical Bayes estimate of the transition rates using the relaxed continuous configuration model recovers the complete network's community structure. Light background areas indicate optimal community assignments. The width of the light blue lines represents teleportation weight. The size of the light blue node centers indicates teleportation probability. The dashed black lines show sample trajectories of random walks. We omit link directions in this example for simplicity.}
    \label{fig:illustration}
\end{figure*}

Instead, we start from the basic idea behind the Bayesian estimate of the map equation and derive an empirical Bayes estimate of the transition rates between nodes in weighted, directed networks. Our Bayesian estimate employs the continuous configuration model \cite{palowitch2018jmlr} and gives a teleportation-like dynamics in a principled way with critical improvements for robust community detection. To ensure an ergodic stationary flow distribution in directed networks, standard teleportation turns a random walker into a random surfer that, besides following links proportional to their weights, teleports uniformly to nodes -- connected or disconnected -- at a fixed rate.
However, teleporting at a fixed rate disregards basic network structure and can wash out significant communities, underfitting the data \cite{lambiotte2012pre}.
Other approaches that reduce the teleportation rate's influence on the community assignments can instead lead to overfitting in networks with missing data.
In our Bayesian estimate of the transition rates, the network flows depend on the amount of available data and network type for robust flow-based community detection in unipartite or bipartite weighted, directed networks with or without metadata (Fig.~\ref{fig:illustration}).

We provide an implementation in Infomap that runs at native speed, available for anyone to download from \href{https://www.mapequation.org}. Using synthetic networks with planted community structures and real-world networks with varying fraction of link observations, we evaluate the empirical Bayes estimate of the transition rates.
We find that Infomap with and without regularized network flows detects similar and robust communities when enough observations are available.
But for incomplete networks with many missing observations, Infomap with empirical Bayes estimates of the transition rates outperforms standard Infomap and prevents spurious communities.

\section{Methodology}

The map equation is an information-theoretic objective function for detecting flow-based communities~\cite{rosvall2008pnas, edler2017algorithms}.
Conceptually, it models network flows as random walks, encodes random walker movements between nodes using codewords, and estimates the theoretical lower limit of the average per-step codelength for a given partition of the nodes into modules.
In line with the minimum description length principle, finding the partition that best compresses the network flows is equivalent to identifying most modular regularities in the network data with respect to those flows.
The Infomap software package \cite{infomap} implements a fast and greedy search algorithm that maximizes flow compression over node partitions by minimizing the map equation.

The basic idea behind the map equation is a communication game where a sender uses codewords to update a receiver about the location of the random walker in the network.
In a one-level partition without modular structure, we assign unique codewords to nodes, and the sender communicates one codeword per random-walker step to the receiver.
The lower limit for the codelength is the Shannon entropy over the nodes' stationary visit rates according to Shannon's source coding theorem \cite{shannon1948mathematical}.
When partitioning nodes into more than one module, we can re-use codewords across modules and achieve shorter average codelengths.
We introduce an index-level codebook to encode transitions between modules and one exit codeword per module for a uniquely decodable code.
The sender uses one codeword to describe transitions within modules and three codewords between modules: for exiting the old module, for entering the new module, and for communicating the visited node in the new module.
In the same fashion, we can extend the coding scheme to hierarchies with three or more levels.
The partition that compresses the flows on the network the most reflects the network's community structure regarding that flow the best.

When sufficiently many observations are available, Infomap returns reliable communities \cite{lancichinetti2009pre, hric2014pre}.
Because the map equation describes the network as-is, missing observations can misrepresent the actual stationary flow distribution, change the balance between module- and index-level codebooks, and distort the communities.
As a consequence, the map equation may capitalize on noise and detect spurious partitions with more and smaller communities than actually present in the complete network \cite{smiljanic2020pre, ghasemian2019overfitting}.

A Bayesian estimate that incorporates prior network assumptions into the map equation overcomes this overfitting problem, and can be derived in closed form for unweighted undirected networks where the stationary visit rate for node $i$ is determined by its degree, $k_i$, as $p_{i} = \frac{k_{i}}{\sum_{i=1}^{N}k_{i}}$ \cite{mitzenmacher2005probability,smiljanic2020pre}.
However, we cannot directly apply this approach to directed or weighted networks for two reasons.
First, we cannot express a corresponding Bayesian estimate of the map equation analytically because no closed-form solution exists for node visit rates in directed networks.
Second, the prior for weighted networks must incorporate link weights absent in previous work \cite{smiljanic2020pre}.
Instead, we formulate an empirical Bayes estimate of a random walker's transition rates to regularize node visit rates \cite{xuanhui2008acm}.

\subsection{The map equation with a Bayesian estimate of the transition rates}

We consider a weighted directed network with $N$ nodes where $A$ represents the adjacency matrix and the matrix $W$ contains information on observed link weights. We assume integer weights for simplicity, but the method also works for non-negative real weights. In general, the probabilities that a random walker steps from node $i$ to other nodes are given by $T_i = (t_{i1}, \dots, t_{iN})$. If we interpret the network as a multigraph, such that $w_{ij}$ denotes the number of observed links between nodes $i$ and $j$, we can explain $W_{i} = (w_{i1}, \dots, w_{iN})$ as a sample of the hidden distribution $T_{i}$. Estimating transition rates $t_{ij}$ using the maximum likelihood estimator gives
\begin{align}
\tilde{t}_{ij} = \frac{w_{ij}}{\sum_{j}w_{ij}}.    
\end{align}
However, with noisy data, $\tilde{t}_{ij}$ can deviate significantly from $t_{ij}$ and cause the map equation to overfit the observed data. To prevent the map equation from overfitting and increase its generalisability, we regularize the transition rates using a Bayesian approach \cite{xuanhui2008acm}. We introduce a prior distribution over $T_i$ and estimate posterior transition rates
\begin{align}
    \hat{t}_{ij}(W_{i}) = \int t_{ij} P(T_i | W_i) dT_i, \label{eq:bayes_estimate}
\end{align} 
where $P(T_i | W_i)$ is a posterior over the unknown distribution $T_i$ given by Bayes' rule,
\begin{align}
    P(T_i | W_i) = \frac{P(W_i | T_i) P(T_i)}{P(W_i)}.
\end{align}
As prior distribution $P(T_i)$, we choose the Dirichlet distribution, which is the conjugate prior of the multinomial distribution and enables analytical calculations:
\begin{align}
    P(T_i | \gamma_i) = \frac{\Gamma(\gamma_{i1} + \dots + \gamma_{iN})}{\Gamma(\gamma_{i1}) \ldots \Gamma(\gamma_{iN})} \prod_{j=1}^{N}t_{ij}^{\gamma_{ij} - 1}.
\end{align}
$\Gamma(x)$ is the gamma function and $\gamma_{i1} \ldots \gamma_{iN}$ are parameters of the distribution.
Given that the likelihood 
\begin{align}
    P(W_i | T_i) = (w_{i1} + \ldots + w_{iN})! \prod_{j=1}^{N} \frac{t_{ij}^{w_{ij}}}{w_{ij}!}
\end{align}
and the total probability of the data
\begin{align}
    P(W_i) = \int P(W_i | T_i) P(T_i) dT_i,
\end{align}
the posterior distribution
\begin{align}
    P(T_i | W_i, \gamma_i) \propto \prod_{j=1}^{N} t_{ij}^{w_{ij} + \gamma_{ij} - 1}. \label{eq:posterior_distribution}
\end{align}
Finally, after integrating Eq.~\ref{eq:bayes_estimate}, we obtain
\begin{align}
    \hat{t}_{ij} &= \frac{w_{ij} + \gamma_{ij}}{\sum_{j=1}^{N}w_{ij} + \gamma_{ij}}\\
                 &= (1 - \alpha_i)\frac{w_{ij}}{\sum_{j}w_{ij}} + \alpha_i \frac{\gamma_{ij}}{\sum_{j}\gamma_{ij}}, \label{eq:bayes_transition_rate}
\end{align}
where $\alpha_i=\frac{\sum_{j=1}^{N}\gamma_{ij}}{\sum_{j=1}^{N}w_{ij} + \gamma_{ij}}$.
The first term is the maximum likelihood estimator weighted by $(1 - \alpha_i)$ and the second term is the transition rates from the prior distribution weighted by $\alpha_i$.
Together they form our empirical Bayes estimate of the transition rates.

The effect of this Bayesian estimate on the transition rates resemble modeling network flows with teleportation. Standard teleportation allows a random walker to teleport uniformly to any node in the network with a fixed small probability $\alpha$ independent of the visited node $i$. Teleportation is necessary to ensure ergodicity in directed networks \cite{brin1998anatomy} but disregards the network structure and turns the flow distribution dependent on the teleportation parameter $\alpha$ \cite{lambiotte2012pre}. For the problem of missing observations, teleportation is not a viable option: For low teleportation rates, the network structure dominates such that the map equation can overfit to noise in the data [Fig.~\ref{fig:illustration}(c)]. Conversely, for high teleportation rates, random jumps dominate over the network structure such that the map equation can underfit and fail to detect relevant community structures.

Interpreting the Bayesian estimate of the transition rates in terms of teleportation, Eq.~(\ref{eq:bayes_transition_rate}) shows that a random walker has node-dependent source and target teleportation probabilities. The random walker chooses an observed link with probability $1-\alpha_i$, or a link in the fully connected prior network with probability $\alpha_i$. In both cases, the probability to follow a link $\left(i,j\right)$ is proportional to its observed weight $w_{ij}$ and prior weight $\gamma_{ij}$, respectively. Thus, if node $i$ has many out-links, the random walker will likely follow them. Otherwise, if the number of out-links of node $i$ is small, it will teleport with a higher probability [Fig.~\ref{fig:illustration}(d)].

How the method performs depends on the parameters $\gamma$. We should choose them such that they can reduce bias induced by incomplete observations while still not wash out regularities in the network structure. We assume that the adjacency matrix $A$ and the weight matrix $W$ are decoupled and use
\begin{align}
    \gamma_{ij} = \lambda_{ij} c_{ij},
\end{align}
where $\lambda_{ij}$ is a connectivity parameter that reflects our prior assumption about connections between nodes $i$ and $j$ and the weight parameter $c_{ij}$ reflects our belief about link weights.

\subsection{The Connectivity parameter}
We use the connectivity parameter $\lambda_{ij} = \lambda = \frac{\ln N}{N}$, which corresponds to the connectivity threshold of random networks.
This $\lambda$-value is the theoretical lower bound on density that guarantees almost surely a giant connected component in the network \cite{erdos1959randomgraph,palasti1966strong}.
When no further node attributes are known, we assume that the connectivity between each pair of nodes is $\lambda = \frac{\ln N}{N}$.
This choice creates a prior network strong enough to prevent overfitting but permissive enough to detect well-supported communities, and works well to regularize the map equation for undirected, unweighted networks \cite{smiljanic2020pre}.
The choice manifests a prior belief that the network is connected without any community structure.
When more information about nodes is available, such as types, classes, or similar, the connectivity parameter, $\lambda_{ij}$, should be adjusted to reflect this information.
We consider two concrete cases, bipartite networks and nodes annotated with metadata.

\subsubsection{Bipartite networks.}
Bipartite networks model interactions between two kinds of node types, $A$ and $B$, where only nodes with different types interact directly.
A connectivity of $\lambda=\frac{\ln N}{N}$ between all pairs of nodes violates the bipartite structure of the network.
To preserve the bipartite nature of the network, we set the connectivity parameter for links between same-type nodes to zero and adjust it for links between different-type nodes.

We assume a bipartite network with $N_A$ nodes of type $A$, $N_B$ nodes of type $B$, and uniform distribution of links between different-type nodes.
As before, we pick the smallest connectivity parameter $\lambda_{AB}$ such that the resulting network is almost surely connected, $\lambda_{AB} = \frac{\ln \left(N_A + N_B\right)}{\min \left(N_A, N_B\right)}$ \cite{saltykov1995discretemath}.
The resulting bipartite prior weight between nodes $i$ and $j$, using bipartite connectivity $\lambda_{AB}$, is
\begin{align}
  \gamma_{ij}^{\text{bi}} = \left(1 - \delta_{t_i t_j}\right) \lambda_{AB} c_{ij},
\end{align}
where $t_i$ and $t_j$ are the types of nodes $i$ and $j$, respectively, and $\delta$ is the Kronecker delta.

\subsubsection{Metadata.}
Real-world networks often contain more information than links.
For example, nodes can have additional metadata.
Metadata have primarily aided in interpreting detected communities.
However, recent studies suggest that complementing network data with metadata for community detection can help overcome limitations and uncertainties in the network structure \cite{yang2013icdm, newman2015natcomm, hric2016prx, emmons2019pre}.

We use discrete metadata to adjust the connectivity parameter.
As before, we connect each pair of nodes uniformly with connectivity $\lambda = \frac{\ln N}{N}$.
In addition, we use the metadata and reinforce connections between nodes with the same label $m$ by $\lambda_m = \frac{\ln N_m}{N_m}$, where $N_m$ is the number of nodes with label $m$.
With metadata labels $m_i$ and $m_j$ for nodes $i$ and $j$, respectively, the adjusted prior link weight is
\begin{align}
  \gamma_{ij}^{\text{meta}} = \left(\lambda + \delta_{m_i m_j} \lambda_{m_i}\right) c_{ij}.
\end{align}

\subsection{Weight parameter}
To incorporate prior assumptions on weights into our method, we use an empirical Bayesian approach~\cite{efron_2010}. An uninformative prior, such as an exponential link weight distribution, is inadequate since it can wash out regularities in the network structure. Instead, we assume that the data carry information about their prior distribution and estimate prior link weights from the networks.

To derive link weights for a prior network, we adapt the so-called continuous configuration model \cite{palowitch2018jmlr}, which estimates the weight of the link from node $i$ to $j$ as
\begin{align}
   c_{ij} = \frac{\sum_{n=1}^{N} k_n^{\text{in}} + k_n^{\text{out}}}{\sum_{n=1}^{N}s_n^{\text{in}} + s_n^{\text{out}}}\frac{s_i^{\text{out}} s_j^{\text{in}}}{k_i^{\text{out}} k_j^{\text{in}}},
 \label{link_weight}
\end{align}
where $k_i^{\text{in}}$ and $k_i^{\text{out}}$ denote observed in- and out-degrees, and $s_i^{\text{in}} = \sum_j w_{ji}$ and $s_i^{\text{out}} = \sum_j w_{ij}$ denote in- and out-strengths for node $i$. The connectivity parameters defined by Eq.~(\ref{link_weight}) preserve expected weights of in- and out- links incident to a node. They provide higher link weights between nodes with strong connections to their neighbors.

This method also works for unweighted and undirected networks.
Undirected networks can be considered as a special case of directed networks where $k_i^{\text out} = k_i^{\text in} = k_i$ and $s_i^{\text out} = s_i^{\text in} = s_i$ for all nodes $i$.
The relaxed continuous configuration model assigns weights $c_{ij} = 1$ to all links for unweighted networks. 
In this case, our method presented here and the Bayesian estimate of the map equation \cite{smiljanic2020pre} provide identical results.
While we can express the effect of the prior network analytically in the Bayesian estimate of the map equation for undirected, unweighted networks, we can also express it as a Bayesian estimate of the transition rates as in Eq.~(\ref{eq:bayes_transition_rate}) and use it with the standard map equation.

We provide an efficient implementation of the Bayesian estimate of the transition rates for anyone to download from \href{https://www.mapequation.org}. The general implementation for regularized network flows works for unipartite and bipartite, unweighted and weighted, undirected and directed networks with and without metadata. The code runs at native speed because it does not express the all-to-all transition rates from the prior distribution in Eq.~(\ref{eq:bayes_transition_rate}) as links.

\section{Results}

We evaluate the performance of Infomap with our empirical Bayes estimate of the transition rates in networks with missing observations.
Our focus is on weighted, directed networks with unweighted and undirected networks as special cases.
For simplicity, we restrict our analyses to networks with integer weights and interpret them as multigraphs, such that link weights $w_{ij}$ denote the number of observed edges between nodes $i$ and $j$.
To create networks with missing observations, we sample from synthetic and empirical multigraphs by removing an $r$-fraction of their multiedges uniformly at random, resulting in reduced edge weights.
For robust results, we average over 100 repetitions for each $r$-value.
As a baseline, we use the performance of the standard map equation and compare the number of detected communities, partition similarity, and predictive accuracy.
We measure partition similarity with the adjusted mutual information (AMI) \cite{vinh2010ami} between detected and planted partition and predictive accuracy with cross-validation. 

\subsection{Synthetic networks}

We use the Lancichinetti-Fortunato-Radicchi (LFR) method \cite{lancichinetti2009pre} to generate a weighted directed network with $N = 1000$ nodes, average node degree $k=7$, and mixing parameter $\eta = 0.4$.
The resulting network has $M=31$ communities and an average link weight of $4.9$ with integer link weights.
We have included results for synthetic networks with different parameters in Appendix~\ref{appendixA}.

To construct synthetic networks with metadata, we first assign metadata labels in perfect alignment with the community assignments of the LFR networks.
Because metadata labels and network community structure are not always aligned \cite{peel2017sciadv}, we assign one of the existing $M = 31$ metadata labels to a $\mu$-fraction of the nodes at random to evaluate the performance for different metadata and community structure correlations.
In this way, we can use the same network to test our empirical Bayes estimate of the transition rates both with and without metadata.

\begin{figure}[ht]
  \includegraphics[width=8.6cm]{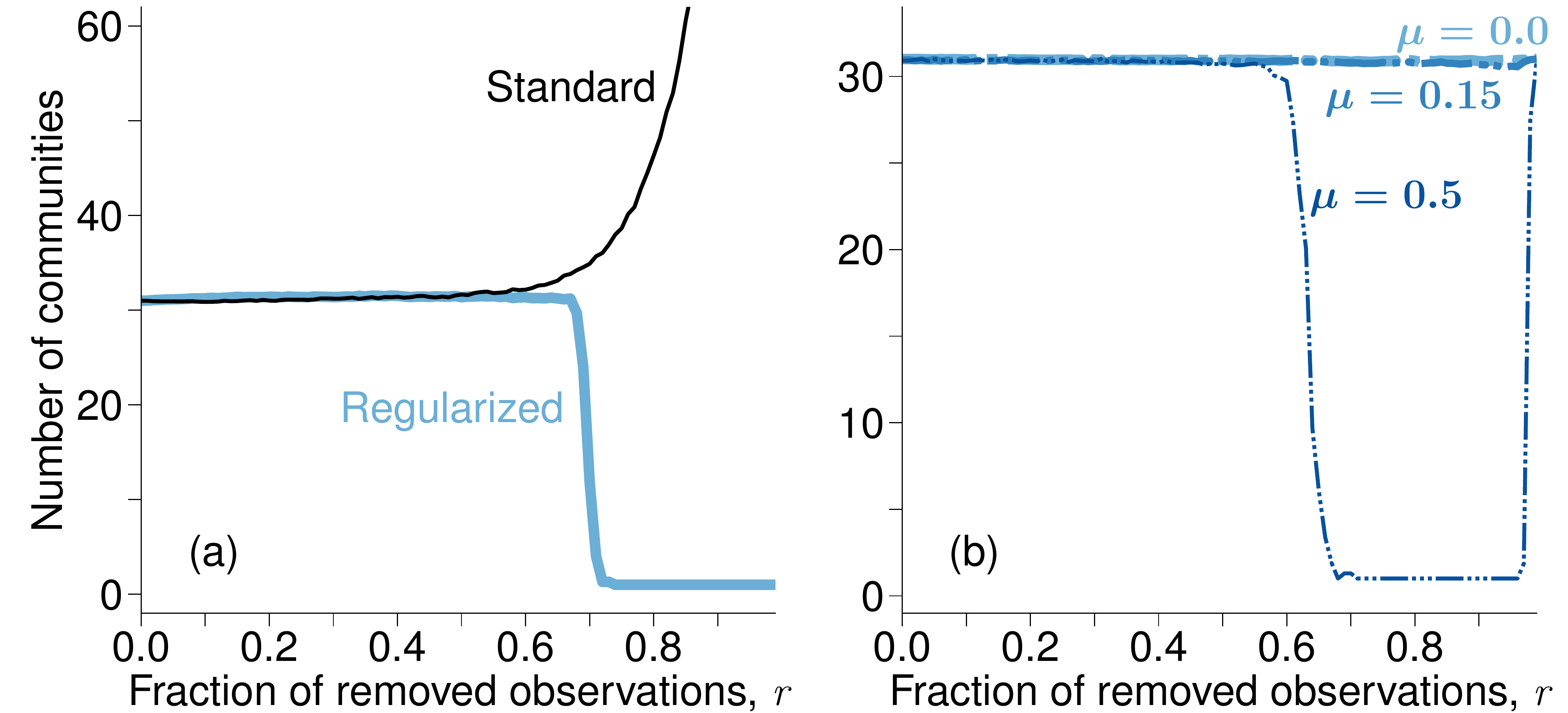}
  \caption{Mean number of communities in synthetic weighted and directed networks with and without our empirical Bayes estimate of transition rates. Without metadata in (a) and with metadata in (b), where a fraction $\mu$ of the nodes have randomly assigned metadata. Results are averages over 100 network samplings.} 
  \label{fig_nummodules_lfr}
\end{figure}

With uniform connectivity and as long as we remove up to half of the edges, corresponding to $r \leq 0.5$, the standard map equation and the map equation with regularized network flows detect virtually the same number of communities [Fig.~\ref{fig_nummodules_lfr}(a)].
When we remove more than half of the data and move beyond $r = 0.5$, the standard map equation begins to detect more and smaller communities. In contrast, the map equation with regularized network flows does not detect community structure anymore.
The relative weight of the prior network increases as we remove more data and the remaining evidence is not strong enough to support communities.

With a metadata-based Bayesian estimate of the transition rates, the fraction of removed links, $r$, does not affect the number of detected communities if the correlation between metadata and planted partition, $\mu$, is high [Fig.~\ref{fig_nummodules_lfr}(b)].
When we randomize half of the metadata labels, corresponding to $\mu = 0.5$, and move beyond the detectability point at $r \approx 0.65$, we find two regimes.
First, two opposing forces are at work, the noisy network structure and the metadata, and we detect no community structure.
Then, as we approach $r = 1$ and almost no link observations remain in the network, we detect the partition corresponding to the metadata labels.

\begin{figure}[ht]
  \includegraphics[width=8.6cm]{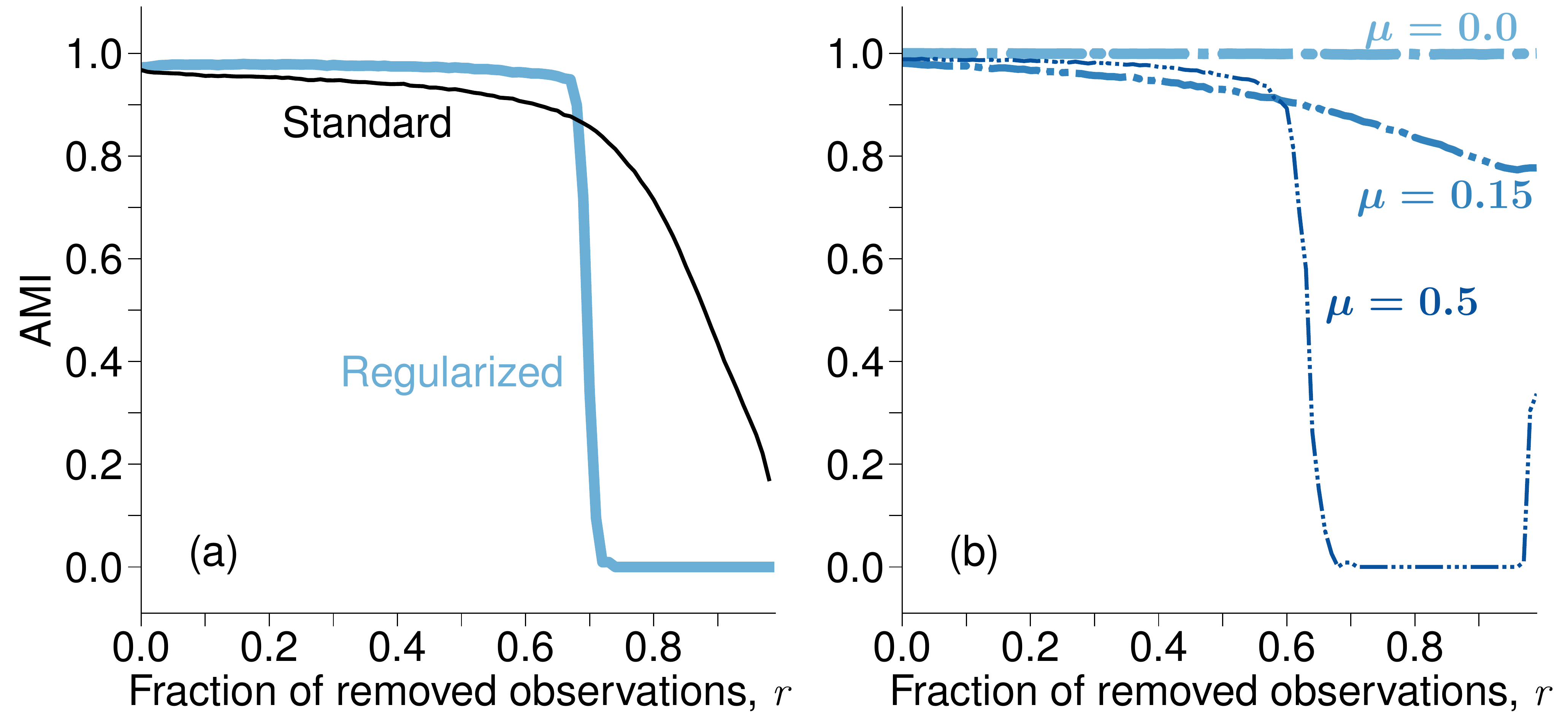}
  \caption{Adjusted mutual information in synthetic weighted and directed networks with and without Bayesian estimate of the transition rates. Without metadata in (a) and with metadata in (b), where a fraction $\mu$ of the nodes have randomly assigned metadata. Results are averages over 100 network samplings.} 
  \label{fig_ami_lfr}
\end{figure}

Although the standard map equation detects the correct number of $M = 31$ communities when we remove less than half of the observations, the AMI scores show that Infomap assigns some nodes to incorrect communities [Fig.~\ref{fig_ami_lfr}(a)].
The map equation with regularized network flows detects communities that better match the planted communities.
When we remove more than half of the observations, $r > 0.5$, the standard map equation detects more communities and the AMI score decreases.
In contrast, the map equation with regularized network flows detects only one community with an AMI score of zero, indicating that the available data is insufficient to infer community structure.

When using a metadata-based Bayesian estimate of the transition rates, our method detects the planted partition reliably if the metadata and the planted partition match perfectly, corresponding to $\mu = 0$ [Fig.~\ref{fig_ami_lfr}(b)].
The method assigns some nodes incorrectly for $\mu > 0$ and weaker correlations with less aligned structural and metadata information.
When many observations are missing, the performance depends on how well the metadata align with the planted community structure.

\begin{figure}[ht]
  \includegraphics[width=8.6cm]{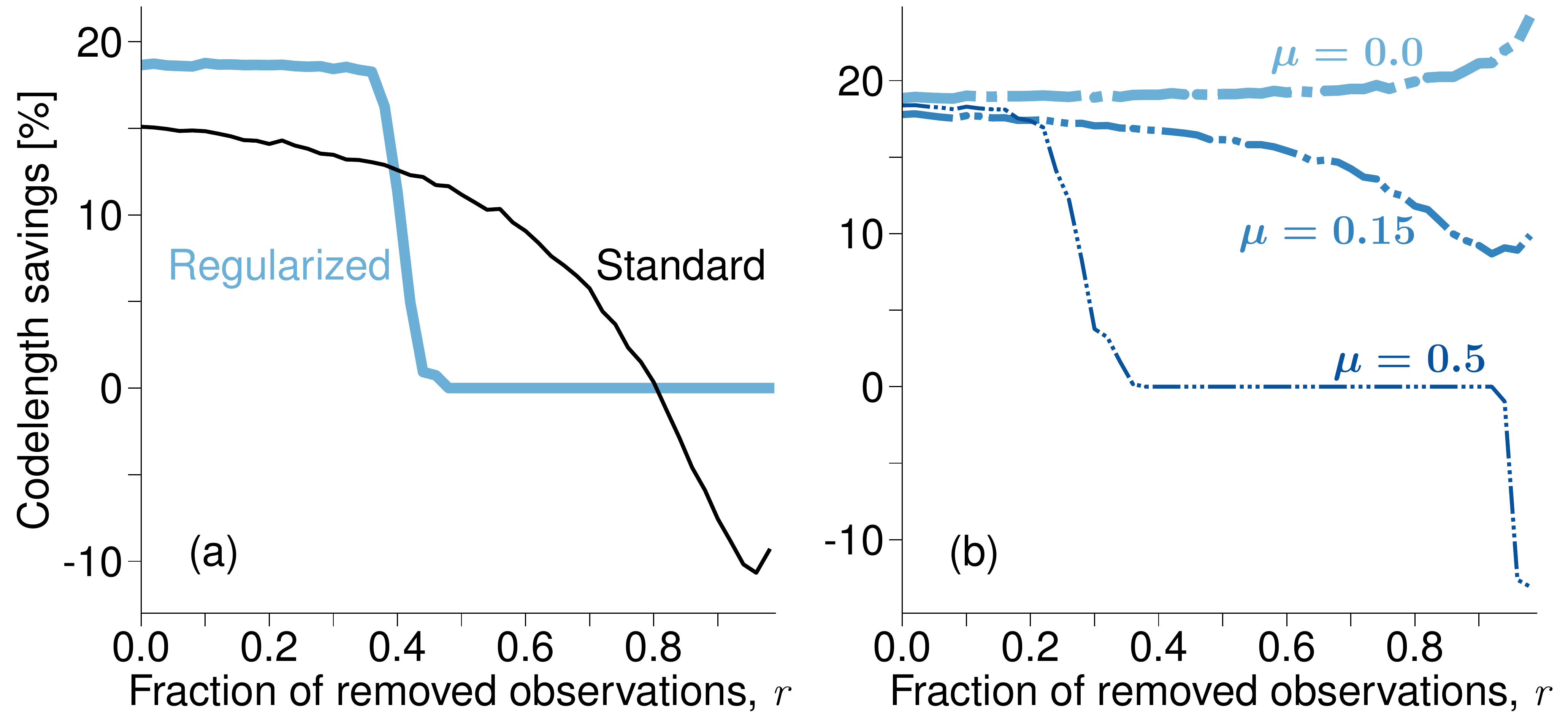}
  \caption{ Codelength savings in synthetic weighted and directed networks with and without regularized network flows. Without metadata in (a) and with metadata in (b), where a fraction $\mu$ of the nodes have randomly assigned metadata. Results are averages over 100 network samplings.} 
  \label{fig_codelength_lfr}
\end{figure}

Many communities and low AMI scores in the undersampled regime indicate that the standard map equation returns spurious communities.
To understand better how this affects the system characterization, we use a cross-validation approach where we first split the multiedge counts of a network into training and test multiedges such that the same edge $\left(i,j\right)$ can occur in the training and validation data, and their counts sum to the original observed count.
Then, we infer the partition that maximizes compression in the training network with Infomap and calculate the test network's description length using that partition.
If the partition captures the structure of the training network well, we expect that it also compresses the description length in the test network.
However, if insufficient data are available in the training network, Infomap overfits and returns a partition that inaccurately describes the structure of the test network, resulting in low compression.
Since the modular description length depends on the number of link observations \cite{smiljanic2020pre}, we construct balanced two-fold splits.
For a multigraph with $m$ observed edges, we choose $\frac{m}{2}$ edges uniformly at random and without replacement for the training network and place the remaining $\frac{m}{2}$ edges in the test network.
Because this split induces further undersampling, we cannot compare the link-removal performance with the previous analysis that started with a complete network. 
Nevertheless, we can use the results to provide more insights into how each method performs in the undersampled regime.

To quantify the level of compression that a partition $\mathsf{M}$ achieves in the test network, we consider the relative codelength savings, the codelength for partition $\mathsf{M}$ compared to the one-module solution $\mathsf{M_1}$ that assigns all nodes to the same module, $l = 1 - \frac{L\left(\mathsf{M}\right)}{L\left(\mathsf{M_1}\right)}$.
Although the standard map equation does not find the optimal partition under incomplete observations, the results indicate that it captures some regularities and achieves positive codelength savings [Fig.~\ref{fig_codelength_lfr}(a)].
However, when the codelength savings are negative, a correct delineation of the network structure is likely infeasible. 
The map equation with regularized network flows and uniform connectivity achieves better compression up until $r \approx 0.4$, indicating that it better captures the network structure.
Beyond this point, and in the regime where the standard map equation detects partitions with negative compression, the map equation with regularized network flows without metadata information assigns all nodes to the same community, resulting in no compression and codelength savings of zero [Fig.~\ref{fig_codelength_lfr}(a)]. 

The map equation with metadata-based Bayesian estimate of the transition rates detects partitions that capture the network regularities well and provide positive codelength savings, even when the metadata labels do not match the planted community assignments for a moderate fraction of the nodes, for example, $\mu = 0.15$. [Fig.~\ref{fig_codelength_lfr}(b)].
However, when the correlation between metadata and planted partition is weak ($\mu=0.5$), and many observations are missing, the method cannot identify significant communities anymore.

\subsection{Empirical networks}

We analyze the performance of the map equation with and without regularized network flows on six empirical networks from different domains where four of the networks are weighted, and three are directed.
\begin{description}
    \item[Sociopatterns] The social network of recorded interactions between female and male students in a high school in Marseille organized as bipartite network \cite{mastrandrea2015plosone}. The students are assigned to one of 9 classes which we use as metadata.
    \item[CoRA] The network covers citations between computer science research papers \cite{macskassy2007jmlr}. The papers are classified into nine different research topics that we use as metadata.
    \item[Industry] The network contains companies that are connected if they appeared together in a business story~\cite{macskassy2007jmlr}. We use Yahoo!'s 12 industry sectors as metadata.
    \item[cit-HepTh] The network contains citations from within arXiv's HEP-TH section \cite{leskovec2005kdd}. We consider only published articles and use information about the journals as metadata.  
    \item[Pok\'emon] Using information from all seven generations of Pok\'emon, we create a network by connecting two Pok\'emon who share the same abilities \cite{pokemon}. The primary type of the Pok\'emon is used as metadata.
    \item[Openflights] The network contains links between non-USA airports \cite{openflights}. We use countries as metadata.
\end{description}
Table~\ref{table:realnet} provides summary information of topological properties of the networks and their metadata. 

\begin{table*}
\caption{Summary of network data. The column Kind denotes if the network is directed (D) or undirected (U). The notations w and M refer to the average link weight and the number of metadata categories in the network, respectively. The last column reports the AMI between metadata and partition detected by the standard map equation in the complete network}
 \label{table:realnet}
 \begin{tabular}
 {Wl{2.5cm}Wr{1.5cm}Wr{1.5cm}Wc{1.5cm}Wc{0.8cm}Wr{0.8cm}Wr{1.0cm}}
 \mytoprule
  Network & Nodes & Links & Kind & w & M & AMI \\
 \mymidrule
  \rule{0pt}{3ex}{Sociopatterns~\cite{mastrandrea2015plosone}} & 143+175 & 2265 & U & 1.33 & 9 & 0.9 \\
 \addlinespace
  {CoRA~\cite{macskassy2007jmlr}} & 3385 & 22092 & D & 1.00 & 9 & 0.3 \\
 \addlinespace
  {Industry~\cite{macskassy2007jmlr}} & 1778 & 14154 & U & 2.79 & 12 & 0.2 \\
 \addlinespace
  {cit-HepTh~\cite{leskovec2005kdd}} & 4378 & 55186 & D & 1.00 & 9 & 0.0 \\
 \addlinespace
  {Pok\'emon~\cite{pokemon}} & 743 & 18184 & U & 1.10 & 18 & 0.3 \\
 \addlinespace
  {Opeflights~\cite{openflights}} & 964 & 8850 & D & 1.48 & 97 & 0.4 \\  
 \mybottomrule
 \end{tabular}
\end{table*}

 \begin{figure}
  \includegraphics[width=8.6cm]{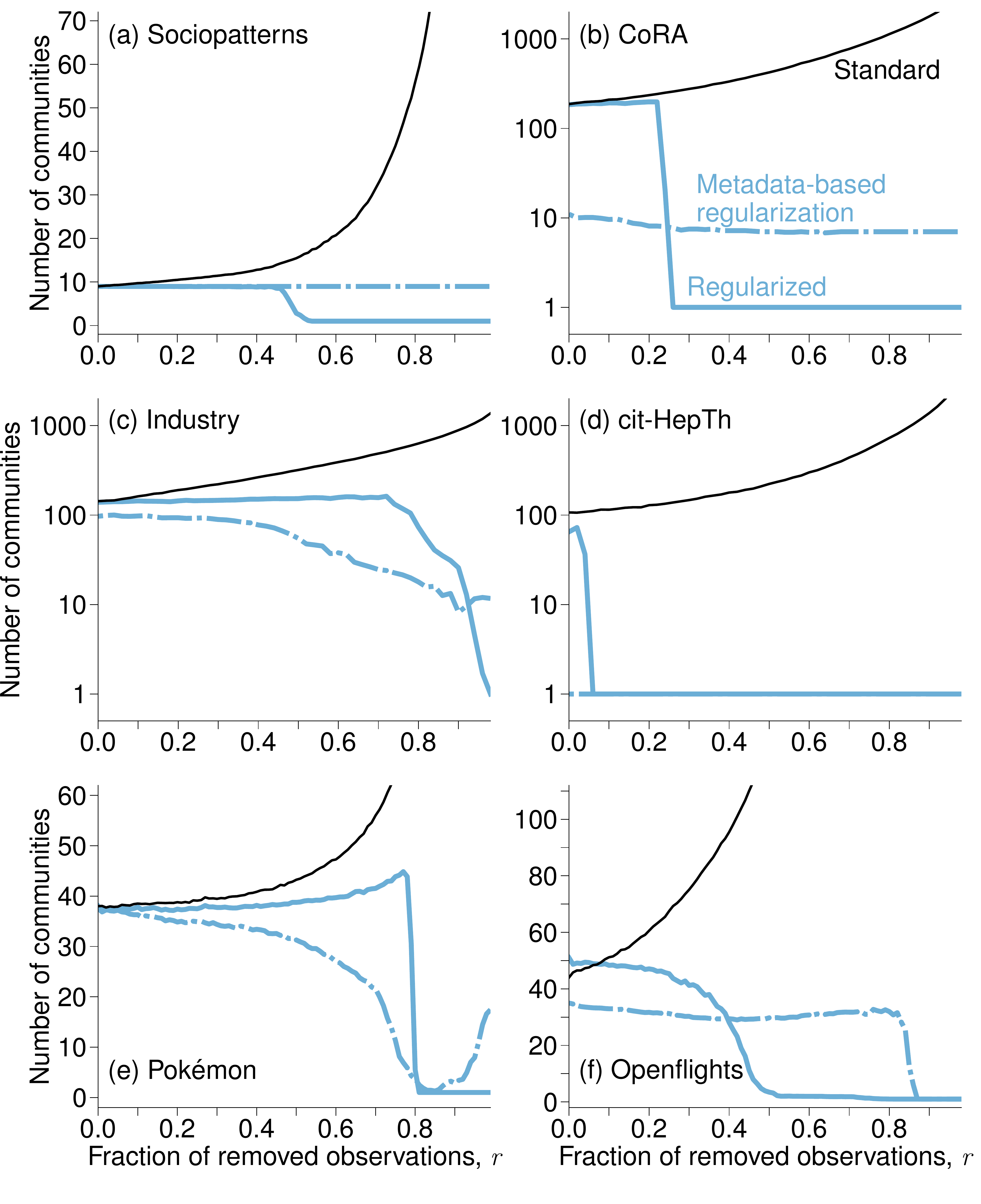}
  \caption{\label{fig_nummodules_realworld} Mean number of communities in empirical networks obtained by the standard map equation, the map equation with teleportation and  uniform connectivity, and the map equation with metadata-based Bayesian estimate of the transition rates. Results are averages over 100 network samplings.} 
 \end{figure}

 \begin{figure}
   \includegraphics[width=8.6cm]{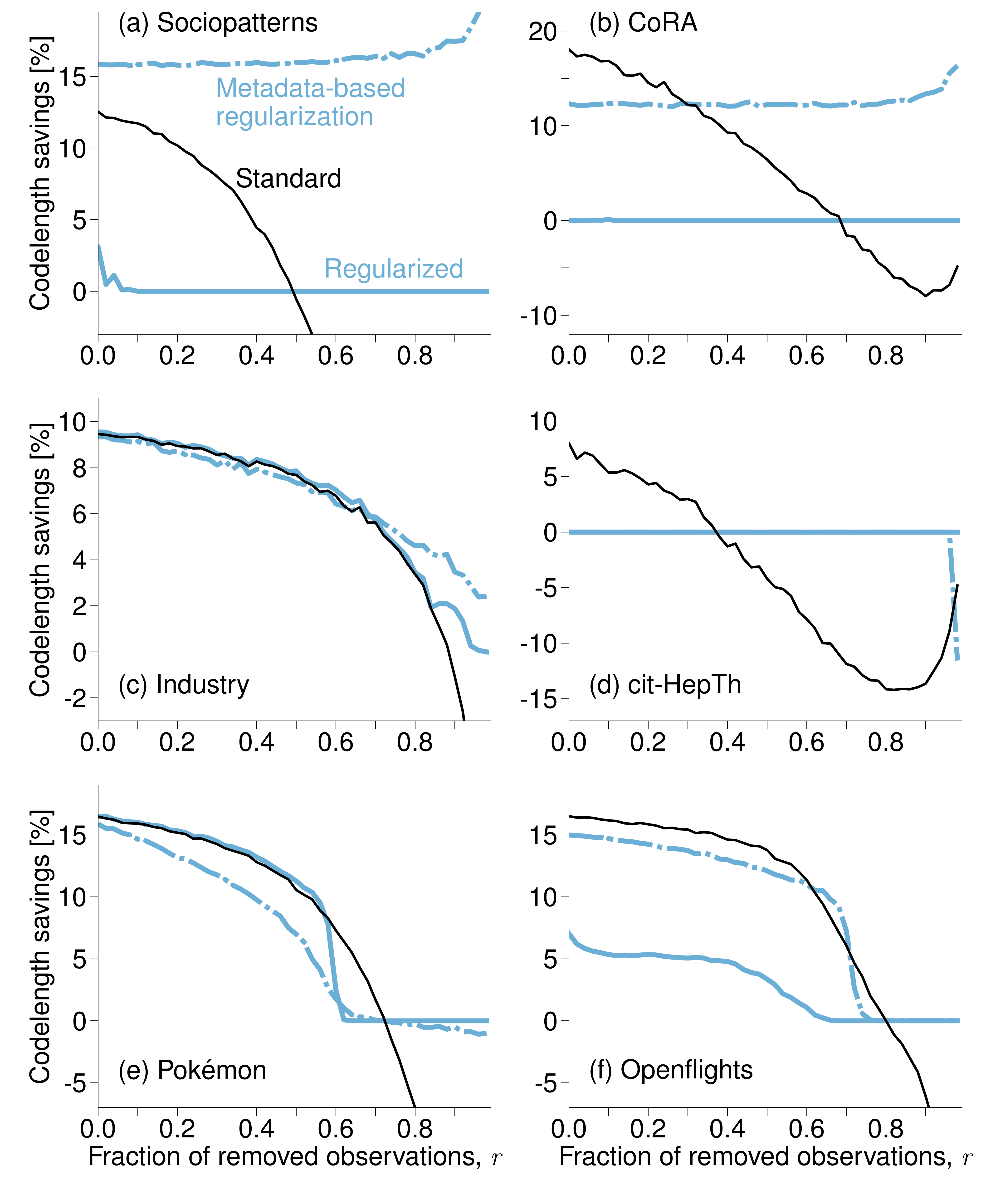}
   \caption{\label{fig_codelength_realworld} Codelength savings in test networks obtained by the standard map equation, the map equation with teleportation and uniform connectivity, and the map equation with metadata-based Bayesian estimate of the transition rates. Results are averages over 100 network samplings.} 
  \end{figure}

We analyze each of the six empirical networks and report the number of communities (Fig.~\ref{fig_nummodules_realworld}) and relative codelength savings (Fig.~\ref{fig_codelength_realworld}).
However, because there is no ground truth partition for empirical data, we cannot use AMI to evaluate our results.

The empirical networks behave like the synthetic networks when analyzed with the standard map equation and the map equation with regularized network flows.
In the complete networks, and when we remove only a small fraction of the observations, the methods detect partitions with a similar number of communities.
When we remove more observations and enter the undersampled regime, the standard map equation detects more and smaller communities.
In contrast, the map equation with regularized network flows without metadata information detects no community structure (Fig.~\ref{fig_nummodules_realworld}).

The empirical networks enter the undersampled regime at different points.
In the Pok\'emon and Industry networks, the map equation with regularized network flows detects communities even after removing $70\%$ of the observations.
In the Pok\'emon network, the number of communities detected by the map equation with regularized network flows increases slightly with the fraction of removed observations before it drops sharply to 1 at $r = 0.8$ and no community structure is detected anymore.
However, the community structure in the cit-HepTh network is sensitive to undersampling, and the map equation with regularized network flows cannot detect communities if we remove more than $~5\%$ of the observations.

The cross-validation results show that partitions with noisy substructures detected by the standard map equation sometimes compress flows on the test network better than the one-level partition.
With more data missing, eventually, the detected partitions lead to negative codelength savings, and the one-level partition offers a better description of the network flows (Fig.~\ref{fig_codelength_realworld}).
The map equation with teleportation does not suffer from this issue.
The mechanism we have implemented prevents overfitting and instead returns the one-level partition when not enough data is available to support community structure in the network.

How well metadata labels align with the network structure determines the performance for the map equation with regularized network flows using metadata.
We use the partitions detected by the standard map equation on the complete networks as a proxy for the network structures and report the AMI with the metadata labels in Table~\ref{table:realnet}.
For example, in the Sociopatterns network, the metadata contains useful information and improves the performance in the undersampled regime.
The number of communities remains the same for all $r$-values [Fig.~\ref{fig_nummodules_realworld}(a)] and, as the cross-validation results show, we achieve high compression in the test network [Fig.~\ref{fig_codelength_realworld}(a)].
In contrast, in the cit-HepTh network, where journals do not support citation patterns between articles, the metadata does not reveal significant communities in the network structure [Fig.~\ref{fig_nummodules_realworld}(d)].
Similarly, in the Pok\'emon network where metadata labels align only weakly with community structure, we observe lower performance than for the map equation with regularized network flows without employing metadata.
When we remove almost all link observations, using uncorrelated metadata can lead to negative codelength savings [Fig.~\ref{fig_codelength_realworld}(d) and (e)].

In the remaining three networks, even though the correlation between metadata and community structure is low, we find that the map equation with regularized network flows benefits from employing the metadata in the undersampled regime.
The map equation with metadata-based Bayesian estimate of the transition rates detects fewer communities than the other two methods.
The higher codelength savings indicate that the detected partitions better capture the structural patterns in the networks by avoiding overfitting to weakly-supported substructures [Fig.~\ref{fig_codelength_realworld}(b),(c), and(f)].

Our analyses show that using regularized network flows with or without metadata prevents overfitting in the undersampled regime.
Instead of returning spurious partitions from sparse observations, the map equation with regularized network flows returns the one-level partition, indicating insufficient evidence to support any community structure.
To detect more regularities with better compression, the metadata-based Bayesian estimate of the transition rates detects more regularities and achieves better compression when correlations between metadata and the network structure are moderate or higher.
With low correlations, the map equation with regularized network flows with metadata can underfit, and the map equation with regularized network flows without employing metadata performs better.

Overall, we recommend the standard map equation for complete network data or when communities from missing links are not problematic. When spurious communities can harm the analysis, the map equation with regularized network flows provides a robust approach. 
 
\section{Conclusion}

We have equipped the flow-based map equation framework with a regulatory mechanism to deal with missing link observations in weighted and directed networks.
By deriving an empirical Bayes estimate of the transition rates that employs a relaxed continuous configuration model, the network flow dynamics account for the uncertainty of observed node degrees and strengths.
The empirical Bayes estimate of the transition rates can incorporate additional information about node types and attributes, enabling extensions to bipartite networks and networks with metadata.
Our adaptable solution also supersedes artificial teleportation for mathematically sound flow modeling on directed networks.

We have implemented the map equation with empirical Bayes estimates of the transition rates in Infomap and analyzed synthetic and real-world networks to evaluate the performance.
Our results show that regularizing the network flows prevents overfitting in undersampled networks, even when a substantial fraction of the data are missing.
Incorporating metadata to reflect prior knowledge about the network can compensate for missing link observations when the metadata correlate with the network structure.
Our results suggest that the map equation with an empirical Bayes estimate of the transition rates provides an effective way to identify robust communities in weighted and directed networks with incomplete observations.

\section*{Code}
We have implemented the map equation with our Bayesian estimate of the transition rates in Infomap. Full documentation of Infomap, including tutorials, instructions and visualization tools is available at \href{https://www.mapequation.org}.

\section*{Funding}
C.B.\ was supported by the Wallenberg AI, Autonomous Systems and Software Program (http://wasp-sweden.org) funded by the Knut and Alice Wallenberg Foundation.
D.E., J.S.\ and M.R.\ were supported by the Swedish Research Council, Grant No.\ 2016-00796.

\appendix

\section{Results for different configurations of synthetic networks}\label{appendixA}

To understand how the the Bayesian estimate of the transition rates affects community detection in networks with different structures, we test the performance on synthetic networks with various sizes, densities, and community strengths.
We create six weighted directed LFR networks with various number of nodes, $N$, average degree, $k$, and mixing parameter, $\eta$, then randomly remove an $r$-fraction of the link observations and detect communities with the standard map equation and the map equation with regularized network flows.

\begin{figure}[ht]
   \includegraphics[width=8.6cm]{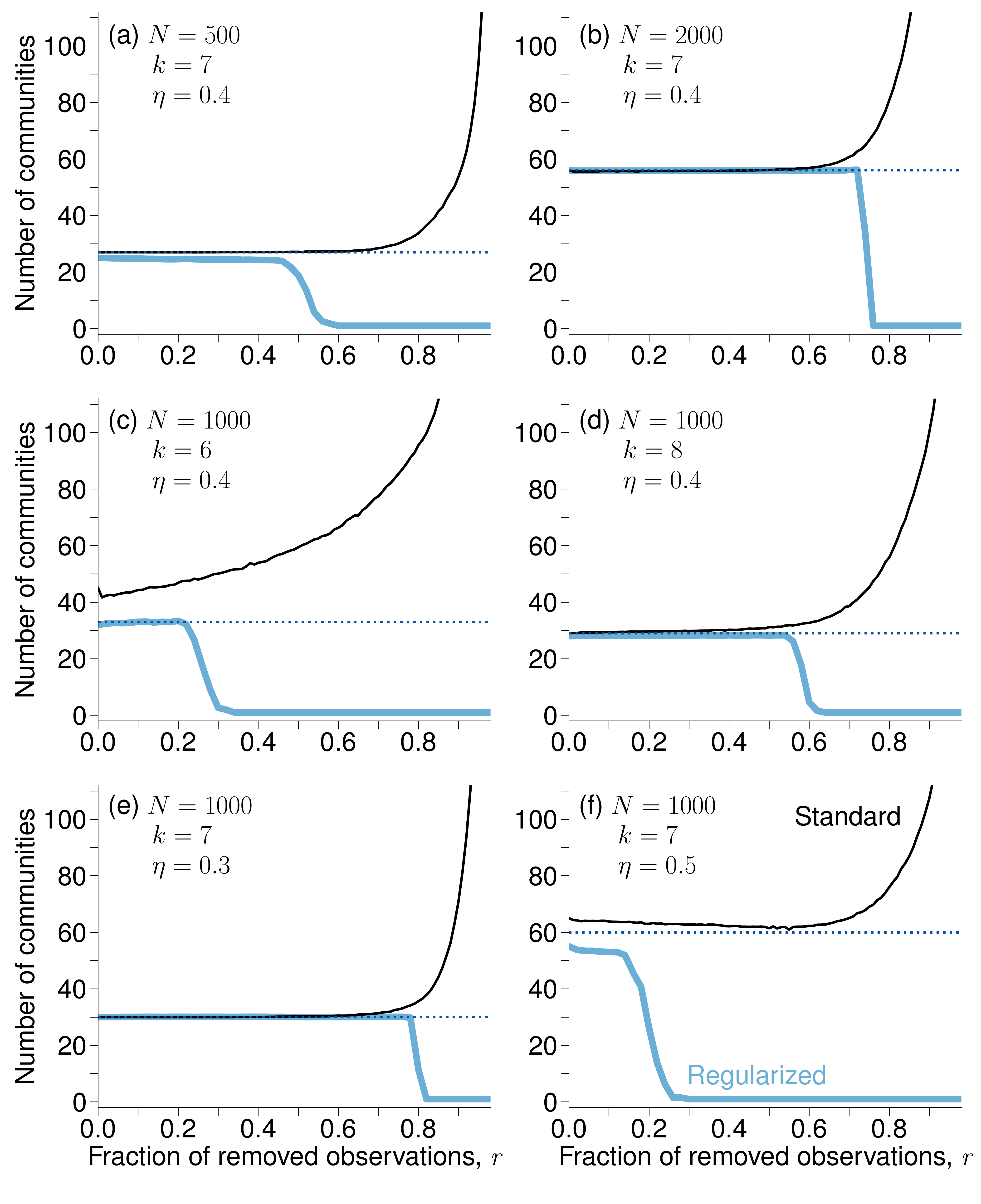}
   \caption{Mean number of communities in synthetic weighted and directed networks with and without regularized network flows. Dotted line indicates number of planted communities.} 
   \label{fig_appendix_nummod} 
  \end{figure}

\begin{figure}
   \includegraphics[width=8.6cm]{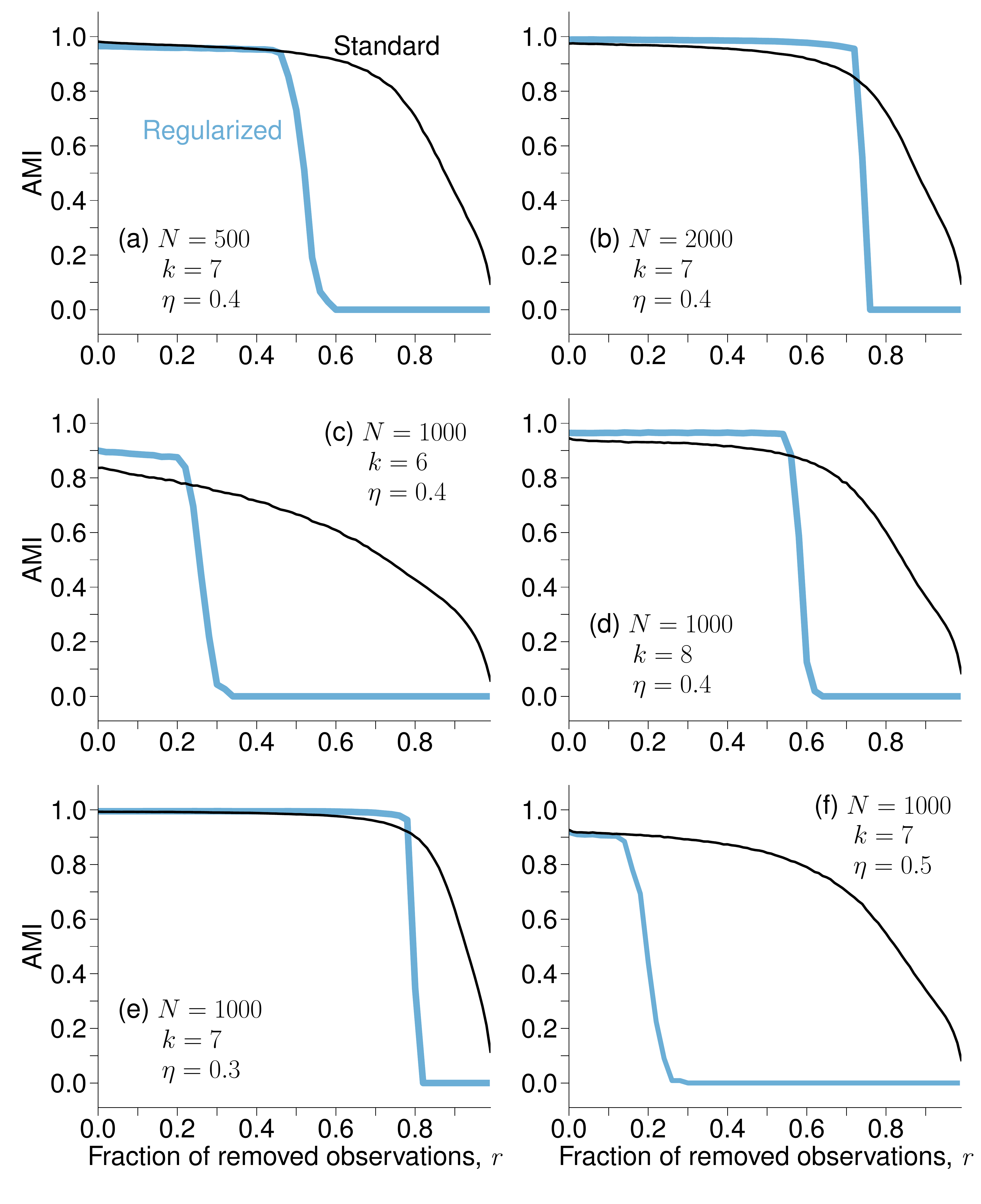}
   \caption{\label{fig_appendix_ami} Adjusted mutual information in synthetic weighted and directed networks with and without regularized network flows.} 
  \end{figure}

Our results show similar trends in terms of robustness to noise in all six networks (Figs.~\ref{fig_appendix_nummod} and~\ref{fig_appendix_ami}).
In the undersampled regime, the performance of the standard map equation decreases fast as the number of missing observations increases.
The map equation with regularized network flows undergoes a sharp transition from detecting robust communities to not detecting any community structure.
The uninformative assumption that a network has no modular structure prevents the map equation with regularized network flows from detecting modular regularities in networks with weak community structure [Fig.~\ref{fig_appendix_ami}(f)].
However, in sparse networks with stronger support for community structure, we find that our Bayesian estimate of the transition rates can improve detection accuracy significantly [Fig.~\ref{fig_appendix_ami}(c)]. 

\newpage

\end{document}